\documentclass[twocolumn,aps,letter]{revtex4}

\usepackage{graphicx,graphics,float}
\usepackage{dcolumn}
\usepackage{bm}
\usepackage{amsmath,amssymb}

\begin{document}

\title{Long Range Free Bosonic Models in Block Decimation Notation: Applications and Entanglement}

\author{Jose Reslen}
\author{Sougato Bose}

\affiliation{Dept. of Physics and Astronomy, University College London, Gower Street, WC1E 6BT London, United Kingdom}%

\date{\today}

\begin{abstract}

We study the effect of long range particle exchange in bosonic arrangements. We show
that by combining the solution of the Heisenberg equations of motion with matrix 
product state representation it is possible to investigate the dynamics as well as 
the ground state while including particle exchange beyond next-neighbours sites. 
These ideas are then applied to study the emergence of entanglement as a result 
of scattering in boson chains. We propose a scheme to generate highly entangled 
multi-particle states that exploits collision as a powerful entangling mechanism.

\end{abstract}

\maketitle

Experimental advancements in low temperature physics increasingly improve the
control and handling of Bose Einstein condensates and other highly correlated 
states of matter 
where quantum mechanics phenomena emerge in their more elemental form.
As a consequence, the gap between experiment and first principle 
analysis via clever numerical methodologies is continuously narrowing down. 
Powerful numerical methods as density matrix renormalization group and time
evolving block decimation (TEBD) \cite{vidal1} allow to simulate systems
of considerable complexity and size, but in most cases both methods 
are limited to short range interactions among neighbour elements. 
Therefore, developing practical methods to carry out
numerics including interactions at long scope is very important, 
as such mechanisms are integral elements of many physical systems, 
e.g., when Coulomb forces are involved. As it is well known, a lot of theoretical
and experimental work has been done seeking to understand quantum systems
and intensive research is currently carried out in many diverse
fields \cite{ResBos,Reslen,Plenio,Rey,Cirac}. In this way motivated, here we
explore practical alternatives for the study of quantum systems. We complement
this discussion by using our findings to explore the
quantum behaviour of 1D boson arrangements. We also focus
on how much entanglement is generated after bosons collide in 
a chain. Although it has been shown that 
interacting wave packets get entangled after collision \cite{Law,Jacksh}, 
most of the studies on this subject focus on
systems made of just two particles, hence applications in many-body 
configurations constitute an important investigation.

Let us consider a 1D arrangement of $N$ sites in which boson exchange can 
take place at all scales. The Hamiltonian is given by

\begin{equation}
\hat{H} = \sum_{k=1}^{N} \sum_{l=k}^N { R_{k,l} \hat{a}^{\dagger}_{k} \hat{a}_l  + h.c} 
\label{eq:one}
\end{equation}

$R_{k,l}$ determines the intensity of the hopping between sites $k$ and $l$ 
while the creation and annihilation operators obey the usual commuting rules
$[\hat{a}_l,\hat{a}^{\dagger}_k]=\delta_k^l$ and 
$[\hat{a}_k,\hat{a}_l]=[\hat{a}^{\dagger}_k,\hat{a}^{\dagger}_l]=0$.
Hamiltonian (\ref{eq:one}) is integrable (quadratic) and the Heisenberg 
equations of motion for the creation operators yield a complete set of 
differential equations of the form 

{\small
\begin{equation}
\frac{d \hat{\alpha}_k}{d t} = -i \sum_{l=0}^N R_{k,l} \hat{\alpha}_l  \Rightarrow \frac{d \mbox{ \boldmath $\hat{\alpha}$ } }{d t} = -i \hat{R} \mbox{\boldmath$ \hat{\alpha}$}, 
\label{eq:two}
\end{equation}
}

subject to the initial condition $\hat{\alpha}_k(t=0)=\hat{a}_k^\dagger$. 
We recall $\hat{\alpha}_k = e^{-i t \hat{H}} \hat{a}^{\dagger}_k e^{i t\hat{H}}$.
The time dependent state ket is given by 
$|\psi(t)\rangle = \prod_{k}^N \hat{\alpha}_k^{n_k} |0 \rangle /\sqrt{\prod_j^N n_j!} $, 
in such a way that the $n_k$'s determine the initial configuration of bosons. 
Similarly, the ground state can be obtained by diagonalizing Hamiltonian 
(\ref{eq:one}). Such ground state can be written as
\vspace{-0.2cm}

\begin{equation}
| G \rangle = \left ( \sum_{k=0}^N c_k \hat{a}^{\dagger}_k \right )^M |0 \rangle,
\label{eq:three}
\end{equation}

where coefficients $c_k$ are the components of the ground eigenvector of matrix $\hat{R}$,
and $M$ is the total number of bosons. While latter expression provides an 
{\it analytical} description of the ground ket, it does not constitute a 
{\it practical} representation of the state for large values of $M$ and $N$.
This holds especially true for computing highly non-local quantities which
require one to deal with the full state. This is because using a local Fock basis 
to write the state demands exponentially growing resources. The purpose of 
this paper is to propose an alternative method to overcome this difficulty. 
Let us operate on the state using unitary operations with the 
intention of simplifying as much as possible Eq. (\ref{eq:three}). First, in order to
make all coefficients real, we operate locally on every site using 
$e^{-i \theta \hat{a}_k^{\dagger} \hat{a}_k}$, where $\theta_k$ represents the 
phase of $c_k$. This produces $\hat{a}_k^{\dagger} \rightarrow e^{-i \theta} \hat{a}_k^{\dagger}$.
 Second, we operate on pairs of consecutive modes,
$\{\hat{a}_{k+1}^{\dagger},\hat{a}_k^{\dagger}\}$, starting with $k=N-1$, using
$e^{-i \phi_k \hat{Q}_k}$, where 
$\hat{Q}_k=\frac{1}{2 i} (\hat{a}_{k+1}^{\dagger} \hat{a}_{k}-\hat{a}_{k}^{\dagger}\hat{a}_{k+1})$.
This induces a rotation-like transformation given by $c_{k+1} \hat{a}_{k+1}^{\dagger} + c_{k} 
\hat{a}_{k}^{\dagger} \rightarrow \left ( c_{k+1} \cos \left( \frac{\phi_k}{2} \right) - c_{k} 
\sin \left ( \frac{\phi_k}{2}  \right )  \right ) \hat{a}_{k+1}^{\dagger} + \left ( c_{k+1} \sin \left( 
\frac{\phi_k}{2} \right) + c_{k} \cos \left ( \frac{\phi_k}{2}  \right )  \right ) \hat{a}_{k}^{\dagger}$.
Hence, operator $\hat{a}_{k+1}^{\dagger}$ can always be cancelled by choosing 
$\tan \left( \frac{\phi_k}{2} \right) = \frac{c_{k+1}}{c_{k}}$. We can carry out
this cancellation consecutively until just the first mode is left operating on
the vacuum, namely,
$|g \rangle = \frac{1}{\sqrt{M!}} \left ( \hat{a}_1^{\dagger} \right )^M |0 \rangle$,
where any overall constant has been absorbed into the definition of $|g \rangle$.
As a result of this decomposition, we can obtain a representation for the ground
state by writing $|g\rangle$ in block decimation notation, also known as matrix
product states (MPS) \cite{Cirac}, which is straightforward,
and then applying the inverse unitary operations in reverse order using the efficient
method of Ref. \cite{vidal1}. Moreover, ground state evolution, for example when
some parameters in the original Hamiltonian are tuned and the whole system
undergoes a quench, can also be studied using the same technique. In such a case
the solution of Eq. (\ref{eq:two}) at any given time is inserted in 
equation (\ref{eq:three}) and the reduction process takes place as above. 
From now on, we will refer to this reduction operation as {\it folding}.

\begin{figure}
\includegraphics[width=0.5\textwidth,angle=-90]{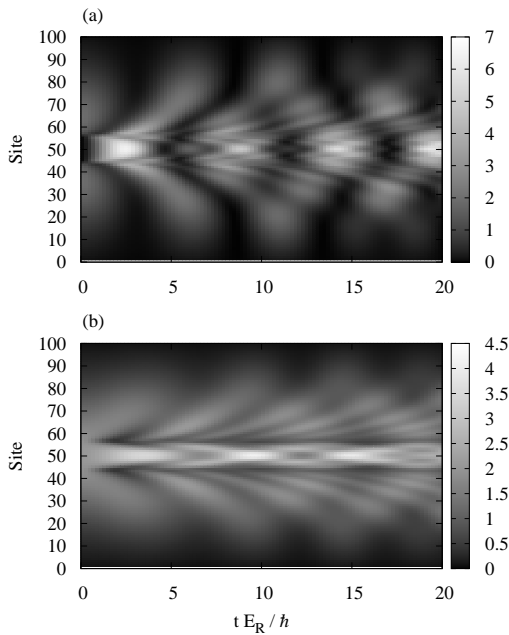}\\
\caption{Average number of bosons. In both simulations the 
hopping among sites decreases inversely with the distance, starting
with $R_{i,i\pm1} = 0.3 E_R$ for first neighbours. $E_R$ is an
energy reference. We
assume that the condensate is trapped in a parabolic potential
with $\Omega=0.00046 E_R$ \cite{Rey}. (a) the initial state corresponds 
to two condensates separated by a potential barrier. This state is 
obtained by calculating the ground state of a chain for the 
specified parameters plus a potential in the middle given by 
$R_{i,i}=1000 E_R$, for $i=45-55$. The dynamics is generated by
turning the barrier off. (b) Starting from the ground state
of the Hamiltonian, a potential barrier like in (a) is turned
on.
\label{fig:1}  
}
\end{figure}

In order to illustrate the method, simulations in chains of 100 
sites and 100 bosons are shown in Fig. \ref{fig:1}. Condensate
dynamics is induced by tuning a potential barrier in the
middle, as experimentally proposed in \cite{Chang}. As a consequence, 
different evolution patterns emerge depending on how and when
the potential barrier is switched on and off. From Fig. \ref{fig:1} we
observe that wave trains travelling outwards are formed moments
after a boson bulk appears in the centre of the chain. Then,
a new bulk reassembles and the process starts again. The wave
patterns generated in this way are consistent with the observations 
reported in \cite{Chang}, where boson trains emerge as a result of the
rich dynamics generated in the experiment. Moreover, the computational
cost involved in the simulations presented above is minimal
compared to TEBD computations, where the chain must be swept
many times until ground state convergence is achieved. 

When several summations of creation operators acting on the vacuum are involved, 
as for instance in systems originally arranged with bosons on different positions, 
state folding can also be implemented. For the shake of simplicity, let us
focus on just two summations so that the state can be written as,

{\small
\begin{equation}
\left ( \overbrace {\sum_{k=1}^N c_k \hat{a}_k^{\dagger}}^{S_2} \right )^{M_2} \left ( \overbrace {\sum_{l=1}^N z_l \hat{a}_l^{\dagger}}^{S_1} \right )^{M_1} |0\rangle,
\end{equation}
}

where $M_{1,2}$ account for the corresponding number of bosons. $S_1$
can be folded using the standard technique, taking into account that the
coefficients in $S_2$ are also affected in the process. After this, we can 
fold $S_2$, but this time folding $\hat{a}_2^{\dagger}$ in $S_2$ would 
inevitably unfold $S_1$. Therefore, the folded state acquires the following
form,

\begin{equation}
|g \rangle = \left( \hat{a}_1^{\dagger} \right )^{M_2} e^{-i \phi_1 \hat{Q}_1} \left (  \hat{a}_1^{\dagger} \right )^{M_1} |0 \rangle.
\label{eq:six}
\end{equation}

In order to write $|g\rangle$ in block decimation notation, we first write 
$|M_1,0,...,0\rangle$
and apply $e^{-i \phi_1 \hat{Q}_1}$. When the state obtained in this way is written with
explicit reference to the local coordinates of the first position, we can apply 
$\left( \hat{a}_1^{\dagger} \right )^{M_2}$ , giving the following result 
\cite{comment1,comment2},

\vspace{-0.2cm}

\begin{equation}
|g\rangle = \sum_{\gamma=1}^{N_\gamma} \Gamma_{1\gamma}^{[1]j}\lambda_\gamma^{[1]} \sqrt{(j+M_2)!} |j + M_2\rangle  |\gamma \rangle^{[2-N]}.
\label{eq:seven}
\end{equation}

From which the canonical coefficients of state $|g\rangle$ can be directly obtained, namely,
${\Gamma'}_{1\gamma}^{[1] j+M_2} = \Gamma_{1\gamma}^{[1]j}$ and ${\lambda'}_\gamma^{[1]} = \sqrt{(j+M_2)!} \lambda_\gamma^{[1]}$
$+$ normalization. It is important to note that this operation, while not unitary, does not
modify the original configuration of Schmidt vectors across the chain. To see this, consider first the
Schmidt vectors corresponding to the first site, $\{|\gamma_i\rangle^{[1]}\}$. Because
Hamiltonian (\ref{eq:one}) preserves the total amount of particles, these vectors 
each have a definite number of bosons associated with them. Consequently, applying $\hat{a}_1^{\dagger}$  
lifts the state occupation by one. These lifted kets are orthogonal and therefore
valid Schmidt vectors of the whole updated state. On the other hand, any other Schmidt 
decomposition of the system is made of only one vector to the left and one vector to the right,
since there is no boson at all between sites $3$ and $N$. As a result, lifting the local basis in
the first site cannot alter the topology of the original decomposition.

In order to illustrate how these ideas can be applied to solve challenging problems, 
we first introduce our dynamical model analytically so as to synthesize 
some useful results. Consider the case where $M$ bosons are sent from one end
of the chain to the other \cite{Plenio,Franco} under the presence of a potential 
barrier in the central part. Evolution is therefore given by 
$\hat{\alpha}_1^M |0\rangle$. We assume that the dynamics is described by a matrix 
$\hat{R} = \hat{J}_x + \epsilon e^{-\beta \hat{J}_z^2}$,
where $\hat{J}_{x,y,z}$ are the standard angular momentum operators.  
In this way, $\hat{J}_x$
induces transmission of bosons from one end of the chain to the other
while the term proportional to $\epsilon$ represents a perturbation 
symmetrically localized over the central part of the chain.
$\hat{R}$ acts on the set of eigenvectors of $\hat{J}_z$, which are associated
to the original operators according to $|j,m \rangle \rightarrow \hat{a}_{j-m+1}^\dagger$, 
where $j=\frac{N-1}{2}$. Similarly, the evolution of operator $\hat{\alpha}_1$ can 
be studied by following the time evolution of ket $|j,j \rangle$:

{\small
\begin{eqnarray}  
&& |\psi(t=\pi) \rangle = e^{-i \pi ( \hat{J}_x + \epsilon e^{-\beta \hat{J}_z^2})}|j,j \rangle \\ 
&& \approx |j,-j \rangle - i \epsilon \int_0^\pi dt e^{-i(\pi-t)\hat{J}_x} e^{-\beta \hat{J}_z^2} e^{-i t \hat{J}_x} |j,j \rangle,  \nonumber
\end{eqnarray}
}

In latter equation we have expanded the time evolution operator to first order
in $\epsilon$. The integral can also be written as

{\small
\begin{eqnarray}
&& \int_0^\pi dt e^{-i(\pi-t)\hat{J}_x} e^{-\beta \hat{J}_z^2} e^{i (\pi-t) \hat{J}_x} e^{-i\pi \hat{J}_x} |j,j \rangle \nonumber \\ 
&& = \int_0^\pi dt e^{-\beta (\cos (t) \hat{J}_z - \sin (t) \hat{J}_y )^2 }|j,-j \rangle \label{eq:tohree}  \\
&& = \frac{1}{\sqrt{4\pi\beta}} \int_{-\infty}^\infty dx e^{-\frac{x^2}{4 \beta}} \int_0^\pi dt e^{i x (cos(t) \hat{J}_z -sin(t)\hat{J}_y )} |j,-j \rangle. \nonumber
\end{eqnarray} 
}

In the last line we have used the well known identity 
$\int_{-\infty}^\infty dx e^{-(a x^2 + b x)} = \sqrt{\frac{\pi}{a}} e^{\frac{b^2}{4 a}}$. 
We can calculate the action of angular momentum operators in the integral above using 
Schwinger's model of angular momentum. To this end, we define two sets of creation and
annihilation operators $\hat{a}_+,\hat{a}^{\dagger}_+$ and $\hat{a}_-,\hat{a}^{\dagger}_-$,
in terms of which 
$\hat{J}_z = \frac{1}{2}(\hat{a}_+^{\dagger} \hat{a}_+ - \hat{a}_-^{\dagger} \hat{a}_-)$,
$\hat{J}_+ = \hat{a}_+^{\dagger} \hat{a}_- $, $\hat{J}_- = \hat{a}_-^{\dagger} \hat{a}_+ $
and $|j,m \rangle = \frac{(\hat{a}_+^{\dagger})^{j+m} (\hat{a}_-^{\dagger})^{j-m}}{\sqrt{(j+m)!(j-m)!}} |0 \rangle $. 
In this way  integral (\ref{eq:tohree}) becomes, 

{\small
\begin{eqnarray}
&& \frac{j}{\sqrt{(2j)!4\pi\beta}} \int_{-\infty}^\infty dy e^{-\frac{j^2 y^2}{4 \beta}}  
\int_0^\pi dt e^{i j^2 x \cos(t)} \times  \label{eq:four} \\
&& \left ( \hat{a}_+^{\dagger} \sin \left ( \frac{j y \sin(t)}{2} \right )  + \hat{a}_-^{\dagger} 
\cos \left (\frac{j y \sin(t)}{2} \right )    \right )^{2 j}  |0 \rangle, \nonumber
\end{eqnarray}
}

where we have performed the variable change $x=jy$. We then use the binomial expansion and 
obtain a sum of integrals. We can expand the trigonometric functions in power series 
with little loss in accuracy when the negative exponential in the first integral
suppresses the contribution of the second integral for relatively high values of $y$. This
can be guaranteed as long as $\beta \ll \frac{j}{2}$, which accounts for a perturbation operating
on a section of the chain much smaller than the system size. With this approximation
equation (\ref{eq:four}) gives,

{\small
\begin{eqnarray}
&& \frac{j}{\sqrt{4\pi\beta}} \int_{-\infty}^\infty dy e^{-\frac{j^2 y^2}{4 \beta}}  
\sum_{k=0}^{2 j} \sqrt{\binom{2 j}{k}} \left (  \frac{y}{2} \right )^{\frac{2j-k}{2}}\times \nonumber \\
&&\Gamma \left ( \frac{2j-k}{2} + \frac{1}{2}\right ) J_{\frac{2j-k}{2}}(j^2 y) |j, j-k \rangle, 
\end{eqnarray}
}

which can be integrated using the asymptotic properties of the Bessel functions. 
So, replacing state vectors by operators we obtain,

\vspace{-0.2cm}

{\small
\begin{eqnarray}
\hat{\alpha}_1(t=\pi) =
\hat{a}_N^\dagger     -i \epsilon \sum_{k=0}^{2 j} \sqrt{\binom{2 j}{k}}
\frac{\Gamma \left ( \frac{2j-k}{2} + \frac{1}{2}\right )^2}
{\Gamma \left ( \frac{2j-k}{2} + 1 \right )} \times 
\beta^{\frac{2 j -k }{2}} \hat{a}_{k+1}^\dagger. \nonumber  
\end{eqnarray}
}

\begin{figure}
\includegraphics[width=0.335\textwidth,angle=-90]{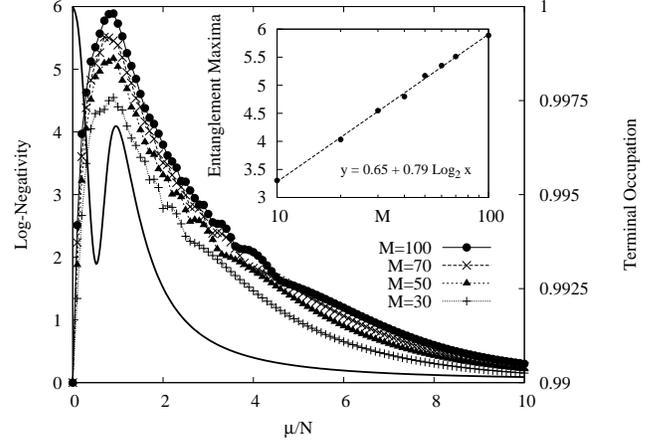}
\caption{\label{fig:2}  
Entanglement between the ends vs. $\frac{\mu}{N}$, the normalized intensity of the 
potential barrier, $\mu = \epsilon e^{-\frac{\beta}{4}}$ (arbitrary energy units),  
in a chain of 100 sites. 
The right axis coordinates correspond to the solid line, the amount of 
bosons on the ends divided by $M$. In this case curves corresponding to
different number of bosons are very similar and therefore they all collapse 
in one single line. Inset. Intensity of entanglement 
maxima in the main figure as a function of the number of bosons.}
\end{figure}

Consequently, for small $\beta$, operator $\hat{\alpha}_1$ will generate
a particle configuration exponentially localized around the end opposite
to the position where particles were initially localized, that is, particles
are efficiently transferred from one end to the other. On the other hand,
for larger $\beta$ the main contribution will come from $\hat{a}_1^\dagger$ 
as well as from $\hat{a}_N^\dagger$. Hence particles will occupy both ends with
little spreading over intermediate sites. Physically, large $\beta$ means 
the perturbation remains tightly localized in the centre of the chain. 
In this case one can think that when the particles get across the centre 
they have a well defined momentum and the perturbation simply acts as a thin 
wall that causes reflection without altering the wave packet 
shape. As a result, particles reflected preserve the coherence necessary 
to be drifted back into their original chain terminal. 
This phenomenon can be used to efficiently generate entanglement among
colliding particles. Ideally, boson packets initially prepared 
far away from each other in a separable state interact in the middle
of the chain via a local potential proportional to the number of boson
on the central positions. Some time after the interaction, the majority 
of bosons turn up on the original positions, but this time these bosons display 
long range entanglement. This process has been simulated using the 
two-sum-folding technique previously introduced. So, we solve equation 
($\ref{eq:two}$) and insert this solution in the expression that determine 
the initial state as previously specified. Then, for a time equal to half a 
period we set the quantum state in MPS representation and find the 
entanglement between the ends of the chain as measured by log-negativity \cite{LogN,ResBos}.
Here we study chains with $N$ even and therefore a potential barrier
is considered in two central sites, but in chains with $N$ odd a highly
localized perturbation can be modelled taking one single central site.
From Fig. \ref{fig:2} it can be seen that particle collection is highly
efficient, with more than $99\%$ of particles collected on the terminals
after the interaction. Similarly, entanglement generation is maximally efficient 
in the range  $0.5<\mu/N<1.5$ (see Fig.\ref{fig:2} caption for the definition of $\mu$)
while the maximum amount of entanglement 
grows logarithmically as a function of the total number of bosons.
This logarithmic behaviour can be studied applying folding techniques. 
To this end, consider the dynamics generated by
$\hat{\alpha}_1^{\frac{M}{2}}\hat{\alpha}_N^{\frac{M}{2}}|0\rangle$.
Because bosons get entangled mainly on the central
part of the chain where the particle packets interact, we focus 
on the form of $\alpha$-operators for a time equal to a quarter period, 
when the boson packets become symmetrically spread along the chain
and therefore $\hat{\alpha}_1(\frac{T}{4}) = \hat{\alpha}_N(\frac{T}{4})$
for a chain without wall. Consequently, the
particle distribution adopts the form of Eq. (\ref{eq:three}), but
this time the the $c_k$'s are determined by the coupling constants and the strength 
of the barrier, both considered general up to symmetry. As entanglement 
between both halves of the chain is independent of unitary transformations
on either half-chain block, we can operate on the state using the 
same transformations employed previously to fold the quantum state as long as we do not
operate on pairs of creation operators belonging to different half-chain blocks.
In this case, the result of the reduction is trivial and can be written as 
$(\hat{a}_1^{\dagger} + \hat{a}_2^{\dagger})^M |0\rangle$. Entanglement can 
be calculated expanding this expression and noticing that as a result we obtain 
a genuine Schmidt decomposition with coefficients $c_k =\sqrt{\frac{1}{2^M}\frac{M!}{(M-k)!k!}}$. 
Hence, log-negativity can be worked out by replacing the binomial distribution
by a normal Gaussian. This gives $E_N \sim \frac{1}{2}log_2 (2 \pi) + \frac{1}{2}log_2 (M)$.
Moreover, assuming that the interaction among bosons takes place in a time scale 
much smaller than the transfer time, we can take this result as an estimation of 
entanglement when the particles reach the chain ends. The discrepancy with the fitting 
constants in Fig. \ref{fig:2} is due to to the approximations involved
in getting $E_N$. 
However, logarithmic growing as well as highly efficient particle 
collection are positively verified by the numerical analysis. Finally, we would 
like to point out that the methodology presented in this work
can be applied to analogous situations in spin chains or fermion arrangements. 
Non linear effects, while less amenable to exact analytical solution, 
could be considered by solving the Gross-Pitaevskii equation and then 
treating the coefficients describing the wave function as the $c_k$'s of 
Eq. (\ref{eq:three}). Additionally, the method can be used
in combination with TEBD, for example in quench-like problems where
repulsion is suddenly turned on. JR acknowledges an EPSRC-DHPA scholarship.


\begin{thebibliography}{}
\bibitem{comment1} In writting $|g \rangle$ in this way, we have used the properties of 
the canonical decomposition as described in Ref. \cite{vidal1}. Specifically, 
we exploit the fact that the elements of tensor $\Gamma_{\alpha \beta}^{[n]i}$ can be
seen as the components of the whole quantum state given in terms of
the local basis, i.e., the occupation number, and the Schmidt vectors
obtained when partitioning the chain to the sides of the place under study.
\bibitem{comment2} In this expression we have implicitly assumed a functional
dependece between $j$ and $\gamma$ due to conservation properties.
See Ref.\cite{ResBos} for a detailed discussion on this matter.
\end{thebibliography}
\end{document}